\documentstyle[12pt]{article}

\font\twelve=cmbx10 at 15pt
\font\ten=cmbx10 at 12pt

\def\Res#1{\raisebox{-1.6ex}{$\stackrel{\textstyle{Res}}{\scriptstyle{#1}}$}}

\begin{document}

\begin{titlepage}

\begin{center}

\renewcommand{\thefootnote}{\fnsymbol{footnote}}

\vspace{3 cm}

{\twelve Pade Approximants and Borel Summation for QCD Perturbation
Expansions}

\vspace{0.9 cm}

\ten { Maciej Pindor 

Institute of Theoretical Physics, Warsaw University

ul. Ho\.za 69

00-681 Warszawa,
Poland}

\vspace{3 cm}

{\ten Abstract}
\end{center}

We study the applicability of Pade Approximants (PA) to estimate a "sum" of
asymptotic series of the type appearing in QCD. We indicate that one should
not expect PA to converge for positive values of the coupling constant and
propose to use PA for the Borel transform of the series. If the latter
has poles on the positive semiaxis, the Borel integral does not
exist, but we point out that the Cauchy pricipal value integral can exist
and that it represents one of the possible "sums" of the original series, the
one that is real on the positive semiaxis. We mention how this method works
for Bjorken sum rule, and study in detail its application to series appearing
for the running coupling constant for the Richardson static QCD potential.
We also indicate that the same method should work if the Borel transform
has branchpoints on the positive semiaxis and support this claim by a simple 
numerical experiment.



\end{titlepage}

\section{Introduction}

Perturbative expansions appearing in QCD and many of its models simmulating
various interesting features of this theory have, usually, coefficients growing
so fast that it is diffcult to obtain an increased precision by naive adding of
subsequent terms \cite{zj},\cite{vz}.

Pade Approximants (PA) are an effective tool to "sum" an asymptotic series 
\cite{bg-m},\cite{gil},
however they may not necesserily sum it to the function that has the required
analytic properties. We must remember that there is an infinite number of 
analytic
functions having the same asymptotic expansion, and the additional requirements
stemming from the physical meaning of the function can help us to diminish this
abundance.

We shall first illustrate this problem on the very simple example and show
that the use of the Borel integral can be of help here.
As we expect
the perturbation series in QCD to have coeffcients growing like $n!K^{n}n^{\gamma}$
\cite{vz} let us consider the simplest series of this type - the Euler series:
\begin{equation} \label{eulers}
e(z)=\sum^{\infty}_{n=0}{n!z^n}
\end{equation}
It is well known \cite{bg-m},\cite{gil} that PA to this series converge 
to the function:
\begin{equation}
E(z)=\int^{\infty}_0{\frac{e^{-u}}{1-zu}du}
\end{equation}
in the whole complex plane with the positive semiaxis excluded. The function 
has the branch point at $z=0$ and PA reconstruct the branch of this function
which is real on the negative semiaxis and has cut along the positive one.
In QCD the expansion parameter is usually the coupling constant and we want 
to have our function for the positive values of coupling constant - moreover
we want, for real expansion coefficients, the function to be real, at least 
close to zero.
Our aim is therefore to find a function which has not only the expansion
(\ref{eulers}) but is also real on the positive semiaxis. We want to stress
that such function is again \underline{not unique}, but we want to show how
one of such functions can be constructed from (\ref{eulers}).

We first consider a case when the Borel transform of the series has only poles
(Section 2) and illustrate our point in Section 3 by two examples of QCD 
related series. Then, in Section 4, we apply our considerations to the case
when the Borel transform has branchpoints and claim that the same approach
(taking the Principal Value of the Borel integral over Pade Approximants to
the Borel transform of the series) would work here - we support this claim
by a simple numerical example.

\section{Poles in the Borel transform}

We consider the series:
\begin{equation} \label{gens}
f(z)=\sum_{n=0}^{\infty}{f_nz^n}
\end{equation}
where we expect that asymptotically (for $n \rightarrow \infty $) $f_n$ behaves like
$n!K^nn^{\gamma}$. The Borel transform of this series is then:
\begin{equation} \label{bors}
g(z)=\sum_{n=0}^{\infty}{g_nz^n}=\sum_{n=0}^{\infty}{\frac{f_n}{n!}z^n}
\end{equation}
If all $f_n$'s are positive (which is often the case in QCD), then we can 
expect
that g(z) has a singularity at $z=K$ (where $K$ is the radius of convergence
of (\ref{bors})). In this case the Borel integral:
\begin{equation} \label{bori}
\Phi(z)=\int_0^{\infty}{e^{-u}g(uz)du}
\end{equation}
is complex for $z>0$, if it exists. If $g(z)(z-K)$ is finite for $z \rightarrow K$,
then this singularity does not spoil the convergence of (\ref{bori}), and 
similiar conditions can be imposed on the behaviour of $g(z)$ near other 
singularities.

Let us first assume that $g(z)$ has a simple pole at $z=K$. In this case:
\begin{equation}
\lim_{z\rightarrow K}g(z)(z-K)=\Res{z=K}\;g(z) < \infty
\end{equation}
and the integral (\ref{bori}) can be written , for $x\epsilon (0,\infty)$ as:

\begin{equation} \label{phip}
\Phi(x)=PV\int_0^{\infty}{e^{-u}g(ux)du} \pm 
                                         \frac{i\pi}{x}e^{-K/x}\Res{x=K}\;g(x)
\end{equation}
- choice of the sign depends on whether in (\ref{bori}) we integrate just below or just
above the positive semiaxis. We see that the imaginary part of $\Phi(z)$ has 
all coefficients of the power expansion at $z=0+$ vanishing and therefore:
\begin{equation} \label{boripv}
{\mathcal{E}}(x)=Re\Phi(x)=PV\int_0^{\infty}{e^{-u}g(ux)du}
\end{equation}
is the function we need for $x>0$.

If we are frustrated by the fact that ${\mathcal{E}}(x)$ is defined only for 
$x>0$, and is not an analytic function in the complex plane, we can define it
also there:
\begin{equation}
{\mathcal{E}}(z)=\int_{0+i\epsilon}^{\infty+i\epsilon}{e^{-u}g(uz)du}+
\frac{i\pi}{z}e^{-K/z}\Res{x=K}\;g(x)
\end{equation}

Obviously we can proceed in an analogous way if $g(z)$ has more poles on
the positive semiaxis. Let us, for the moment, assume it has only poles there.

We now ask how can we construct a sequence of approximations to 
${\mathcal{E}}(z)$ from (\ref{gens}). We have already pointed out that the
direct application of PA to (\ref{gens}) will most probably fail, because
we can expect that PA will diverge on the positive semiaxis. However, as
we know that PA are particularily efficient in approximating functions in
domains of their meromorphy \cite{Nut}, we can use PA to approximate 
$g(z)$ from 
(\ref{bors}), i.e. we propose to construct $[M/N]_{g}(z)$ and approximate
${\mathcal{E}}(z)$ by:
\begin{equation} \label{boripa}
{\mathcal{E}}_{M,N}(z)=PV\int_0^{\infty}{e^{-u}[M/N]_{g}(uz)du}
\end{equation}
The convergence of PA in measure in the neighbourhood of the positive semiaxis
(where we assumed $g(z)$ to be meromorphic, and therefore we expect this type
of convergence)  
guarantees us that (\ref{boripa}) will converge to (\ref{boripv}).
We want to point out that the idea of combining PA with the Borel method was
proposed quite a time ago \cite{marz1},\cite{marz2}, but in a context where
the Borel transform was regular on the positive semiaxis. Essentially the same
concept was used in \cite{pr}, where the Partial Pade Approximant \cite{ppa}
was used to predict a sum of the series for the Gell-Mann-Low function of the 
effective charge.

Before we consider a more general situation when g(z) has also branch points
on the positive semiaxis, let us see how the method works for two examples
from QCD.

\section{Two QCD series with only poles on the positive semiaxis}

First, we consider the Bjorken sum rule. It is known \cite{lt} that in the 
large-$\beta_0$
approximation it has only four simple poles in the Borel plane and therefore
our prescription would be exact for all $[M/N]$ with $M\leq3$ and $N\leq4$.
For realistic values of $\beta_0$, the perturbation series has been calculated
up to 3rd order and there exists an estimate for the next coefficient 
\cite{EGKS}.
It has been demonstrated in \cite{EGKS} that [2/1] PA to the Borel transform
of the truncated series for $f(x)$ (we take $N_f=3$):
\begin{equation}
f(x)=1-x-3.58x^2-20.22x^3-130x^4
\end{equation}
where f(x)is defined in:
\begin{equation}
\int_0^{\infty}{[g^p_1(x,Q^2)-g^n_1(x,Q^2)]dx}=
\frac{1}{6}\mid g_A \mid f(\alpha_s/\pi)
\end{equation}
exhibits a pole at $y=1.05$ 
($y=\frac{\beta_0}{4}\frac{\alpha_s}{\pi}$), 
while we expect poles at $\pm1,\pm2$ in the 
large-$\beta_0$ limit, as mentioned above. Therefore we believe that
\begin{equation}
f(z)=1-zPV\int_0^{\infty}{e^{-u}[2/1]_S(u\frac{\beta_0}{4}z)du}
\end{equation}
is a reasonable approximation to $f(z)$ (for the exact definition of $S$
see \cite{EGKS}). We shall not discuss this example in more depth, but 
only remark that the value of $\alpha_s(Q^2=3GeV^2)$ found this way is 
$.371^{+.075}_{-.068}$.

The next example, which we want to discuss in detail, is the asymptotic 
series for the running coupling constant obtained from the static QCD 
potential for which we take the Richardson potential \cite{jez6}.
We follow notations and conventions of \cite{jez}, and for more clarity
we shall cite some formulae from this paper.

The Richardson potential in momentum space is:
\begin{equation}
V_R(q)=-\frac{16\pi^2C_F}{\beta_0q^2ln(1+q^2/\Lambda^2_R)}
\end{equation}
and we use $n_f=3$, therefore $\beta_0=11-2/3n_f=9$, also $\Lambda_R=.4$
and $C_F=4/3$.
This example is very instructive because one can find the exact formula for
the perturbative coupling in the position space:
\begin{equation} \label{alphaex}
\bar{\alpha}_R(1/r)=\frac{2\pi}{\beta_0}\left[1-\int_1^{\infty}{\frac{du}{u}
\frac{e^{-\Lambda_Rru}}{ln^2(u^2-1)+\pi^2}}\right]
\end{equation}
At the same time the formula for the coupling in momentum space is:
\begin{equation}
\alpha_R(q)=\frac{4\pi}{\beta_0}\left[\frac{1}{ln(1+q^2/\Lambda^2_R)}-
\frac{\Lambda^2_R}{q^2}\right]
\end{equation}
From this, one can calculate the $\beta$ function:
\begin{equation}  \label{beta}
\beta_R(\alpha_R)=q^2\frac{\partial\alpha_R(q)}{\partial q^2}=-\frac{\beta_0}
{4\pi}\alpha^2_R+\frac{\beta_0}{4\pi}\frac{\left[\frac{\beta_0}{4\pi}-
\alpha_R\right]^2}{1+q^2/\Lambda^2_R}
\end{equation}
On the other hand, one has (for any potential) the expansion:
\begin{equation}
\bar{\alpha}_V(1/r)=\sum_{n=0}^{\infty}f_n\left(-\beta_V(\alpha_V)
\frac{\partial}{\partial \alpha_V}\right)^n \alpha_V(q=1/r')
\end{equation}
where $r'=re^{\gamma_E}$, and $\gamma_E$ is Euler's constant.
$f_n$'s are coefficients of the expansion of:
\begin{equation}  \label{fzeta}
f(u)=\sqrt{\frac{tan(\pi u)}{\pi u}}exp\left[\sum_{n=1}^{\infty}
\frac{(2u)^{2n+1}}{2n+1}\zeta(2n+1)\right]=\sum_{n=0}^{\infty}{f_nu^n}
\end{equation}
For details see \cite{jez}.

For later considerations it will be useful to observe that $f(u)$ can also be
written in a different form:
\begin{equation} \label{fmer}
f(u)=\frac{1}{1-2u}\prod_{k=1}^{\infty}\frac{1+\frac{u}{k}}{1-\frac{2u}{2k+1}}
\end{equation}
From this formula we can immediately see that f(u) has zeros at negative
integer values of u and simple poles at half-integer positive values.

It is important to observe from (\ref{beta}) that $\beta_R(\alpha_R)$ is not
analytic at $\alpha_R(q)=0$. The reason is that $\alpha_R(q) \rightarrow 0$
for $q\rightarrow \infty$ and therefore:
\begin{equation}
\beta_R(\alpha_R)=-\frac{\beta_0}{4\pi}\alpha^2_R\ +\quad terms \ like \quad
e^{-\frac{4\pi}{\beta_0\alpha_R}}
\end{equation}
Summing up, we see that we can get for $\bar{\alpha}_R(1/r)$ the perturbation
expansion when we neglect "higher twists":
\begin{equation} \label{alphas}
\bar{\alpha}^{pert}_R(1/r)=\sum_{n=0}^{\infty}{\alpha_R(q)\left(
\frac{\beta_0\alpha_R(q)}{4\pi}\right)^n n!f_n}
\end{equation}
with $q=1/r'$.
The direct calculation of PAs to this series confirms our expectations:
the majority of zeros and poles of diagonal PAs $[M/M]$ (the same is the
case for paradiagonal sequences of the type
$\nolinebreak{[M\pm1/M]}$) lie on the positive semiaxis and interlace:

\vspace{2mm}
\begin{tabular}{rrrrr}
&\multicolumn{2}{c}{zeros}&
\multicolumn{2}{c}{poles} \cr
[2/2]&-155.24210&&-1.91389&\cr
&0.17087&&0.15702&\cr
[3/3]&0.03832&+0.22543i&0.03913&+0.23920i\cr
&0.03832&--0.22543i&0.03913&--0.23920i\cr
&0.09784&&.09658&\cr
[4/4]&-0.14019&0.22494i&-0.14597&0.21387i\cr
&-0.14019&-0.22494i&-0.14597&-0.21387i\cr
&0.15711&&0.14879&\cr
&0.05773&&.05770&\cr
[5/5]&-0.26941& 0.20414i&-0.24492& 0.18468i\cr
&     -0.26941&-0.20414i&-0.24492&-0.18468i\cr
&0.20733&&0.18936&\cr
&0.07728&&0.07709&\cr
&0.0403998&&0.0403993&\cr
[6/6]&-0.32614& 0.18282i&-0.27820& 0.17139i\cr
     &-0.32614&-0.18282i&-0.27820&-0.17139i\cr
&0.23994&&0.21487&\cr
&0.09094&&0.09047&\cr
&0.05022&&0.05021&\cr
&0.03037432&&0.03037431&\cr
[7/7]&-0.32989& 0.18119i&-0.28017& 0.17065i\cr
     &-0.32989&-0.18119i&-0.28017&-0.17065i\cr
&0.24307&&0.21733&\cr
&0.09256&&0.09205&\cr
&0.05154&&0.05153&\cr
&0.03165505&&0.03165504&\cr
&0.01674128&&0.01674128&\cr
\end{tabular}

We see that for large $q$ ($\alpha_R$ small), there will be difficulties 
in finding stable values for PAs, because of condensation of their zeros
and poles.

When we now go to the Borel plane we get the series which is just:
\begin{equation}
\alpha_R(q)f(\frac{\beta_0\alpha_R(q)}{4\pi})
\end{equation}
However we know already (see (\ref{fmer})) that f(u) has only poles
on the positive semiaxis, and therefore we can try to sum (\ref{alphas})
using the formula:
\begin{equation} \label{alphab}
\left . \bar{\alpha}_R(1/r)_{M,N}= \frac{1}{z}PV\int_0^{\infty}
{e^{-u/z}[M/N]_f(u)du}
\right |_{z=\frac{\beta_0\alpha_R(q)}{4\pi}}
\end{equation}

\begin{tabular}{rrrrr}
&\multicolumn{2}{c}{zeros}&
\multicolumn{2}{c}{poles} \cr\cr
[2/2]&0.42052& 0.50473i&0.66625&\cr
&0.42052&-0.50473i&2.23334&\cr
[3/3]&0.95277&-0.74029i&-0.78648&-0.77908i\cr
&0.95277& 0.74029i&-0.78648& 0.77908i\cr
&-1.66887&&0.50170&\cr
[4/4]&-0.94714&&-1.20630&\cr
& 1.00530&-1.10274i&-0.90978&-1.09780i\cr
& 1.00530& 1.10274i&-0.90978& 1.09780i\cr
&2.33029&&0.49992&\cr
[5/5]&-0.99840&&-2.50005&-0.89026i\cr
& 0.73068& 1.56833i&-2.50005& 0.89026i\cr
& 0.73068&-1.56833i&-1.35858& 1.40476i\cr
& 1.27993& 0.84019i&-1.35858& 1.40476i\cr
& 1.27993&-0.84019i& 0.499997&\cr
[6/6]&-1.00023&&-1.72341& 0.82573i\cr
& 0.81310& 1.67371i&-1.72341&-0.82573i\cr
& 0.81310&-1.67371i&-1.16083& 1.62300i\cr
& 1.08549&&-1.16083&-1.62300i\cr
& 1.33197& 0.92750i& 0.50000&\cr
& 1.33197&-0.92750i& 1.07610&\cr
[7/7]&-0.99994&&-1.77158& 0.96646i\cr
& 0.74517& 1.93873i&-1.77158&-0.96646i\cr
& 0.74517&-1.93873i&-1.24094& 1.79463i\cr
& 1.31201& 1.27438i&-1.24094&-1.79463i\cr
& 1.31201&-1.27438i& 0.50000&\cr
& 1.59142& 0.56621i& 1.60358&\cr
& 1.59142&-0.56621i&76.62431&\cr
[10/10]&-1.95927&&-2.25821&\cr
&-1.00000&&-2.14754& 1.01218i\cr
& 0.91881& 2.53060i&-2.14754&-1.01218i\cr
& 0.91881&-2.53060i&-1.77230& 1.74984i\cr
& 1.54963& 1.87556i&-1.77230&-1.74984i\cr
& 1.54963&-1.87556i&-1.16222& 2.47111i\cr
& 1.95905& 1.23582i&-1.16222&-2.47111i\cr
& 1.95905&-1.23582i& 0.50000&\cr
& 2.24957& 0.55669i& 1.50020&\cr
& 2.24957&-0.55669i& 3.90255&\cr
\end{tabular}

\vspace{5mm}
We see that PAs to $f(u)$ actually reconstruct zeros and poles on the real
axis and that there are many other complex zeros and poles drifting away
from $u=0$, most probably "simulating" the essential singularity of $f(u)$
at $u=\infty$.

Now we can see what values for (\ref{alphas}) can one get from (\ref{alphab}).
We shall compare these values with the ones obtained in \cite{jez} as the best
partial sums (denoted by $\bar{\alpha}_{R,N}(1/r)$), 
and those calculated from formula (\ref{alphaex}), though one
must keep in mind, that (\ref{alphaex}) gives the exact value of $\alpha_R$
in position space, while we are here estimating the sum (\ref{alphas}) which
is (\ref{alphaex}) "minus higher twists".

We present below the case $M=N$, but include also $[2/3]$, as it is the lowest
PA which gives already values quite close to the limit and uses only five 
coeffcients of the expansion.
\vspace{3mm}
$$\begin{array}{cllll}
1/r[GeV]&
\multicolumn{1}{c}{10}&
\multicolumn{1}{c}{20}&
\multicolumn{1}{c}{50}&
\multicolumn{1}{c}{100}\cr\cr
[2/2]&.2903148980&.2370475373&.1814661174&.1522256606\cr
[2/3]&.2710750452&.2232228306&.1757241919&.1497191873\cr
[3/3]&.2714196149&.2235681574&.1759161959&.1498234840\cr
[4/4]&.2714564564&.2234913019&.1758530906&.1497850862\cr
[5/5]&.2714137805&.2234817444&.1758530534&.1497859540\cr
[6/6]&.2714097805&.2234792391&.1758723578&.1497857339\cr
[7/7]&.2713988440&.2234777131&.1758524051&.1497857952\cr
[8/8]&.2714007078&.2234779029&.1758523985&.1497857906\cr
\bar{\alpha}_R(1/r)&.2714046&.2218345&.1746993&.1491174\cr
\bar{\alpha}_{R,N}(1/r)&.2856&.22262&.17647&.14868\cr
\end{array}$$

\vspace{3mm}
$$\begin{array}{cllll}
1/r[GeV]&
\multicolumn{1}{c}{10^3}&
\multicolumn{1}{c}{10^4}&
\multicolumn{1}{c}{10^5}&
\multicolumn{1}{c}{10^6}\cr\cr
[2/2]&.09908701930&.07409640462&.05937036386&.04957967229\cr
[2/3]&.09908856884&.07413913414&.05938321829&.04958332635\cr
[3/3]&.09909570614&.07413847053&.05938281763&.04958320088\cr
[4/4]&.09909193228&.07413842675&.05938289532&.04958322765\cr
[5/5]&.09909214599&.07413844576&.05938289564&.04958322732\cr
[6/6]&.09909214737&.07413844662&.05938289573&.04958322732\cr
[7/7]&.09909214937&.07413844659&.05938289572&.04958322732\cr
[8/8]&.09909214928&.07413844654&.05938289571&.04958322732\cr
\bar{\alpha}_R(1/r)&.09901686&.07413084&.05938213&.04958315\cr
\bar{\alpha}_{R,N}(1/r)&.098937&.07413536&.05938222&.04958339\cr
\end{array}$$

\vspace{3mm}
We see that $\bar{\alpha}_R(1/r)_{M,M}$ converge when M grows, but not
to the values of $\bar{\alpha}_R$ - we attribute this to the difference made
by "higher twists". In other words we claim that the procedure we propose
sums the series (\ref{alphas}) well, which allows us to see clearly the 
difference between the perturbative solution - represented by this series -
and the full solution (\ref{alphaex}) including also nonperturbative 
effects. We demonstrate, therefore, once again the fact that the perturbation
series contains only a part of the information about the full solution and
that the requirement that the solution must be real for positive values
of the coupling constant is not sufficient to compensate for the information
contained in "higher twists" terms.

\section{Branch points in the Borel transform}

If the Borel transform has a branchpoint at $z=K$, instead of a pole,
then the integral (\ref{bori}) for $x\epsilon (0,\infty)$ instead of 
(\ref{phip}) takes the form:

\begin{equation} \label{phib}
\Phi(x)=\int_0^{\infty}{e^{-u}Reg(ux)du}\pm 
                                           i\int_{K/x}^{\infty}{e^{-u}Img(ux)du}
\end{equation}
\vspace{3mm}
if $g(z)(z-K)$ is finite for $z\rightarrow K$ (what is a condition for
the existence of (\ref{bori}).
Again, we see that the second integral, which is equal to the imaginary part
of $\Phi(x)$, has all terms of the asymptotic expansion for $x=0+$ vanishing,
and therefore it is the first integral which is of interest for us.

In the same way, as in the case of the simple pole, we can find an integral 
representation for the analytic function having the required asymptotic 
expansion (\ref{gens}) and being real on the positive semiaxis:

\begin{equation}
{\mathcal{E}}(z)=\int_{0+i\epsilon}^{\infty+i\epsilon}{e^{-u}g(uz)du}-
     \frac{1}{2z}\int_K^{\infty}{e^{-u/z}[g(u+i\epsilon)-g(u-i\epsilon)]du}
\end{equation}
\vspace{3mm}
These facts were also observed earlier \cite{Grun}, but we want to stress that 
$Re\Phi(x)$ has the same asymptotic expansion at zero as $\Phi(x)$ and 
therefore it cannot be reliably estimated by calculation of partial sums.

If we want now to use the same concept as before: to sum (\ref{bors}) using
PAs, we must remember that they will most probably not converge on semiaxis 
$x\epsilon (K,\infty)$. We conjecture, however, that if PAs to g(z) converge
in measure arbitrarily close to positive semiaxis (where we expect a cut,
"reconstructed" by PAs, should lie), then the integral:

\begin{equation} \label{boribap}
{\mathcal{E}}(z)_{M,N}=PV\int_0^{\infty}{e^{-u}[M/N]_g(uz)du}
\end{equation}

will converge to $Re\Phi (z)$ (see (\ref{phib})) for z>0.

To check a plausibility of this hypothesis, we try to calculate this way:

\begin{equation}
F(x)=Re\int_0^{\infty}{e^{-u/x}log(1-u)du}
\end{equation}

i.e. we approximate $F(x)$ by:

\begin{equation}
F_{M,N}(x)=PV\int_0^{\infty}{e^{-u/x}u[M/N]_{log(1-t)/t}(u)du}
\end{equation}

As $log(1-t)/t$ is the Stieltjes function, all poles of PAs to it lie on
the semiaxis $(1,\infty)$ and interlace with zeros \cite{bg-m},\cite{gil}.

On the other hand, $F(x)$ can easily be converted to a simple principal
value integral:

\begin{equation} \label{blex}
F(x)=-xPV\int_0^{\infty}{\frac{e^{-u/x}}{1-u}du}
\end{equation}

We give below values of $F_{M,N}(x)$ for few values of $x$ and compare 
them with values obtained from a numerical integration of the last formula.

$$\begin{array}{crrrrr}
x&\multicolumn{1}{c}{.05}&
\multicolumn{1}{c}{.1}&
\multicolumn{1}{c}{.2}&
\multicolumn{1}{c}{.95}&
\multicolumn{1}{c}{5}\cr
&&&&&\cr
[1/1]& 0.0527903605& 0.1128659958& 0.2754741870& 0.4092156955& 0.049324951\cr
[1/2]& 0.0527972269& 0.1131155255& 0.2731042382& 0.8349862840&-1.544989018\cr
[2/2]& 0.0527977128& 0.1131547428& 0.2705616695& 0.7678390728&-1.080166535\cr
[2/3]& 0.0527977828& 0.1131540942& 0.2703267712& 0.6479952192&-0.590917162\cr
[3/3]& 0.0527977927& 0.1131486878& 0.2708042139& 0.7437635764&-1.199222773\cr
[3/4]& 0.0527977949& 0.1131464797& 0.2708502503& 0.7091822199&-0.256031716\cr
[4/4]& 0.0527977953& 0.1131466717& 0.2707469311& 0.6971934572&-0.961775112\cr
[5/5]& 0.0529797953& 0.1131470772& 0.2707743167& 0.7066673691&-0.711782921\cr
[6/6]& 0.0527977953& 0.1131470109& 0.2707637347& 0.6701782542&-0.586264408\cr
[7/7]& 0.0527977953& 0.1131470222& 0.2707667730& 0.7125737482&-0.577295644\cr
[8/8]& 0.0527977953& 0.1131470201& 0.2707662425& 0.7110753552&-0.621454057\cr
[9/9]& 0.0527977953& 0.1131470205& 0.2707662164& 0.7110508353&-0.665707635\cr
[10/10]& 0.0527977953& 0.1131470205& 0.2707662706& 0.7113684213&-0.688595754\cr
[11/11]& 0.0527977953& 0.1131470205& 0.2707662537& 0.7114624672&-0.691851149
\end{array}$$

It sholud be noticed that starting from $[8/8]$ the quadruple precision was 
necessary to obtain the above numbers.
\section{Conclusions}

We have proposed above to use PAs to sum the Borel transforms of the series
appearing in QCD, and then to calculate Cauchy Principal Value of the Borel 
integral to obtain a function which has the given series as its asymptotic
expansion at zero, and which is at the same time real on the positive 
semiaxis in the complex coupling constant plane. The proposal is not entirely
new \cite{pr}, but we have pointed out that it produces values of the well
defined analytic function, one of the variety of them having the same 
asymtotic expansion. We have also argued that the
method should work not only in the case when the Borel transform has poles
on the positive semiaxis, but also when it has branchpoints there.
This latter point is illustrated by the very simple numerical experiment.

\section{Acknowledgements}

The author is indebted to prof. M. Je\.zabek for drawing his attention to his 
paper \cite{jez}, and for interesting discussion and also to dr. P. R\c aczka
for pointing him out paper \cite{pr}.

\end{document}